\documentclass[12pt]{article}
\usepackage[latin9]{inputenc}
\usepackage{geometry}
\usepackage{bm}
\usepackage{amsmath}
\usepackage{amssymb}
\usepackage{graphicx}

\usepackage{babel}
\begin{document}
\title{Rare decays of the positronium ion and molecule, $\text{Ps}^{-}\to e^{-}\gamma$
and $\text{Ps}_{2}\to e^{+}e^{-}\gamma,~\gamma\gamma,~e^{+}e^{-}$}
\author{M.~Jamil Aslam$^{\text{(a,b)}}$, Wen Chen$^{\text{(a)}}$, Andrzej
  Czarnecki$^{\text{(a)}}$, \\ 
  Samiur Rahman Mir$^{\text{(a)}}$
 , and
Muhammad  Mubasher\textsuperscript{(a)}
\\
\textit{\small{}$^{\text{(a)}}$Department of Physics, University
of Alberta, Edmonton, Alberta, Canada T6G 2E1}\\
\textit{\small{}$^{\text{(b)}}$Department of Physics, Quaid-i-Azam
University, Islamabad, Pakistan}}
\date{\vspace*{-65mm}\hfill Alberta Thy 26-21 \vspace*{60mm}}

\maketitle
\global\long\def\me{m}%

\global\long\def\calA{\mathcal{A}}%
\global\long\def\calE{\mathcal{E}}%
\global\long\def\calL{\mathcal{L}}%
\global\long\def\calM{\mathcal{M}}%

\global\long\def\va{\bm{a}}%
\global\long\def\vb{\bm{b}}%
\global\long\def\vj{\bm{j}}%
\global\long\def\vk{\bm{k}}%
\global\long\def\vm{\bm{m}}%
\global\long\def\vn{\bm{n}}%
\global\long\def\vp{\bm{p}}%
\global\long\def\vq{\bm{q}}%
\global\long\def\vr{\bm{r}}%
\global\long\def\vs{\bm{s}}%
\global\long\def\vu{\bm{u}}%
\global\long\def\vv{\bm{v}}%
\global\long\def\vw{\bm{w}}%
\global\long\def\vx{\bm{x}}%
\global\long\def\vy{\bm{y}}%
\global\long\def\vz{\bm{z}}%

\global\long\def\vA{\bm{A}}%
\global\long\def\vB{\bm{B}}%
\global\long\def\vD{\bm{D}}%
\global\long\def\vE{\bm{E}}%
\global\long\def\vF{\bm{F}}%
\global\long\def\vH{\bm{H}}%
\global\long\def\vJ{\bm{J}}%
\global\long\def\vK{\bm{K}}%
\global\long\def\vL{\bm{L}}%
\global\long\def\vM{\bm{M}}%
\global\long\def\vN{\bm{N}}%
\global\long\def\vP{\bm{P}}%
\global\long\def\vQ{\bm{Q}}%
\global\long\def\vR{\bm{R}}%
\global\long\def\vS{\bm{S}}%
\global\long\def\vU{\bm{U}}%
\global\long\def\vX{\bm{X}}%

\global\long\def\val{\bm{\alpha}}%
\global\long\def\vrho{\bm{\rho}}%
\global\long\def\vgammma{\bm{\gamma}}%
\global\long\def\vom{\bm{\omega}}%
\global\long\def\vga{\bm{\gamma}}%
\global\long\def\vep{\bm{\epsilon}}%
\global\long\def\vnabla{\bm{\nabla}}%
\global\long\def\vmu{\bm{\mu}}%
\global\long\def\vnu{\bm{\nu}}%
\global\long\def\vsi{\bm{\sigma}}%
\global\long\def\vsigma{\bm{\sigma}}%
\global\long\def\vSi{\bm{\Sigma}}%

\global\long\def\order#1{\mathcal{O}\left(#1\right)}%

\global\long\def\edge#1{\left.#1\right|}%
\global\long\def\d{\mbox{d}}%

\global\long\def\bra#1{\left\langle #1\right|}%
\global\long\def\ket#1{\left| #1 \right\rangle }%

\global\long\def\G{\widetilde{G}}%

\global\long\def\tr{\mbox{Tr}}%

\global\long\def\Li{\mbox{Li}_{2}}%

\global\long\def\az{\alpha_{Z}}%

\global\long\def\ap{\alpha_{\pi}}%

\global\long\def\za{Z\alpha}%

\global\long\def\Ep{E_{\bm{p}}}%

\global\long\def\Eb#1{E_{{\scriptscriptstyle \text{bind},#1}}}%

\global\long\def\bs{\blacksquare}%

\global\long\def\GF{G_{{\scriptscriptstyle \text{F}}}}%

\begin{abstract}
Decay rates of the positronium molecule $\text{Ps}_{2}$ into two
photons and into an electron-positron pair are determined. Previous
studies find that these rates are very different, 
\[
\Gamma\left(\text{Ps}_{2}\to e^{+}e^{-}\right)/\Gamma\left(\text{Ps}_{2}\to\gamma\gamma\right)\simeq250\text{ (previous studies)}.
\]
This is puzzling since both processes have two body final states and
are of the same order in the fine structure constant. We propose a
simple calculational method and test it with the well-established
decay of the positronium ion into an electron and a photon. We then
employ it to correct predictions for both these $\text{Ps}_{2}$
decays.  We find that previous studies overestimated the $e^{+}e^{-}$
and underestimated the $\gamma\gamma$ channel by factors of about 5.44
and 3.93 respectively. Our results give
$\Gamma\left(\text{Ps}_{2}\to
  e^{+}e^{-}\right)/\Gamma\left(\text{Ps}_{2} \to \gamma\gamma\right)
\simeq 11.7$.
\end{abstract}

\section{Introduction}

The lightest bound state involving an electron and a positron is positronium,
$\text{Ps}$. Its ground state is the spin-singlet para-positronium.
The spin triplet is called ortho-positronium. In quantum electrodynamics
(QED), due to the charge conjugation invariance, para- and ortho-positronium
can decay only into an even and odd numbers of photons, respectively.
At least two photons must be produced because of momentum conservation.

Positronium with an additional electron or positron forms a positronium
ion $\text{Ps}^{\pm}$, first observed in 1981 \cite{Mills:1981zzc}.
Very efficient methods of the ion production have recently been
developed  \cite{PsIonSource08,Nagashima201495}.
The extra constituent makes one-photon annihilation, $\text{Ps}^{\pm}\to e^{\pm}\gamma$,
possible \cite{PhysRev.44.510.2}. First theoretical studies of this
decay in Ps$^{-}$ \cite{Misawa:1985dd,Ho_1983} and in Ps$^{+}$
\cite{Chu:1986zza} were incomplete and were corrected by Kryuchkov
\cite{Kryuchkov_1994} who included all contributing Feynman diagrams.
As a warmup for our main calculation, we confirm and simplify Kryuchkov's
analysis. 

Two positronium atoms can form a molecule, $\text{Ps}_{2}$, first
considered by Wheeler in his seminal study of compounds of electrons
and positrons which he called polyelectrons \cite{Wheeler:1946xth}.
Its binding energy was first computed by Hylleraas and Ore \cite{Hylleraas:1947zza}.
70 years after that theoretical demonstration $\text{Ps}_{2}$ was
discovered by Cassidy and Mills \cite{Cassidy:2007wq}. Various properties
of $\text{Ps}_{2}$ including the precise binding energy of its ground
and excited states as well as rates of major and some minor decay
modes have been established in a number of papers, including \cite{Frolov:1995zz,Frolov:1997hy,Varga:1998ss,Puchalski:2007ck,Puchalski:2008jj,Frolov:2009qi}.

In the ground state of $\text{Ps}_{2}$ electrons and positrons both
form spin singlets, a feature important for this paper. This is energetically
favorable because the antisymmetry, necessary for identical fermions,
originates in the spin configuration. The spatial wave function is
symmetric under the exchange of electron coordinates (similarly for
the positrons), and therefore less curved, minimizing the kinetic
energy.

Electrons' spins are uncorrelated with those of the positrons. A random
encounter of an electron with a positron can therefore result in an
annihilation into an even or an odd number of photons. Typically,
only one $e^{+}e^{-}$ pair annihilates and the remaining $e^{+}e^{-}$
constituents are liberated. Such processes usually produce only two
photons but higher numbers are also possible, just like in atomic
positronium decays \cite{Frolov:2009qi,Czarnecki:1999mt}.

In addition, more than two constituents can interact in the decay
process. Such reactions are rare because $\text{Ps}_{2}$ is weakly
bound and inter-particle distances are large, on the order of the
Bohr radius{{} $a_{B}=1/\alpha\me$, where $\alpha\simeq1/137$
is the fine structure constant and $\me$ is the electron mass. Annihilation
involves virtual particles whose typical propagation range is the
electron Compton wavelength, suppressed by an additional factor $\alpha$.
When an electron and a positron meet, the probability that there are
$n$ additional constituents within a Compton distance scales approximately
like $\alpha^{3n}$. }

Despite this huge suppression, we find these rare decays theoretically
interesting. $\text{Ps}_{2}$ is the simplest known four-body bound
state and serves as a model for more complicated systems such as tetraquarks
\cite{Czarnecki:2017vco,Karliner:2017qjm,Ali:2009es,Hernandez:2019eox}.
In principle, all properties of this molecule can be calculated with
arbitrary precision within QED. However, this few-body system is sufficiently
intricate that even some of its tree-level decays have not yet been
correctly evaluated. 

In this paper, we focus on two decays that involve all four constituents:
$\text{Ps}_{2}\to e^{+}e^{-}$ and $\text{Ps}_{2}\to\gamma\gamma$.
The rate of the radiationless decay $\text{Ps}_{2}\to e^{+}e^{-}$
was first studied in Ref.~\cite{Frolov:1995zz}, subsequently re-derived
and confirmed in \cite{bailey2005} and further refined in \cite{Frolov:2009qi},
\begin{equation}
\Gamma\left(\text{Ps}_{2}\to e^{+}e^{-};\text{{Ref.\ \cite{Frolov:2009qi}}}\right)=2.3\cdot10^{-9}\text{ s}^{-1}.
\end{equation}
The rate of the so-called total annihilation, $\text{Ps}_{2}\to\gamma\gamma$,
was calculated more recently \cite{Perez-Rios:2014qia},
\begin{equation}
\Gamma\left(\text{Ps}_{2}\to\gamma\gamma;\text{{Ref.\ \cite{Perez-Rios:2014qia}}}\right)=9.0\cdot10^{-12}\text{ s}^{-1}.
\end{equation}
The very different magnitudes of these rates contradict intuitive
arguments presented above. Both are two-body decays involving all
four constituents of $\text{Ps}_{2}$ and occurring in the same order
in $\alpha$. Why do their rates differ by a large factor? Using published
formulas \cite{Frolov:2009qi,Perez-Rios:2014qia} one finds
\begin{equation}
\frac{\Gamma\left(\text{Ps}_{2}\to e^{+}e^{-};\text{{Ref.\ \cite{Frolov:2009qi}}}\right)}{\Gamma\left(\text{Ps}_{2}\to\gamma\gamma;\text{{Ref.\ \cite{Perez-Rios:2014qia}}}\right)}=\frac{512}{521}\cdot147\sqrt{3}\simeq250.\label{eq:250}
\end{equation}
This is the puzzle we set out to clarify. In Sec.~\ref{sec:Single-photon-decay}
we propose a simple approach to calculating decay amplitudes of polyelectrons
such as $\text{Ps}^{-}$ and $\text{Ps}_{2}$. We test it with the
example of the positronium ion decay $\text{Ps}^{-}\to e^{-}\gamma$
and find agreement with Ref.~\cite{Kryuchkov_1994}. We also confirm
the rate of an analogous process in the molecule, $\text{Ps}_{2}\to e^{+}e^{-}\gamma$,
previously published in \cite{Frolov:2009qi}. In Sec.~\ref{sec:Two-photon-annihilation}
we apply this technique to determine $\Gamma\left(\text{Ps}_{2}\to\gamma\gamma\right)$,
and in Sec.~\ref{sec:ee-annihilation} we present our result for
$\Gamma\left(\text{Ps}_{2}\to e^{+}e^{-}\right)$. We find the ratio
of these rates to be 
\begin{equation}
\frac{\Gamma\left(\text{Ps}_{2}\to e^{+}e^{-}\right)}{\Gamma\left(\text{Ps}_{2}\to\gamma\gamma\right)}=\frac{27\sqrt{3}}{4}=11.7.
\end{equation}

We conclude in Sec.~\ref{sec:Conclusion} with comments on the magnitude
of our result and with an attempt to clarify what went wrong in the
previous studies \cite{bailey2005,Perez-Rios:2014qia,Frolov:2009qi,Frolov:1995zz}.
Appendices present spinor configurations and symmetry factors for
the $\text{Ps}^{-}$ and $\text{Ps}_{2}$ decay amplitudes.

\section{Three-constituent annihilation process $e^{+}e^{-}e^{\pm}\to e^{\pm}\gamma$\label{sec:Single-photon-decay}}

In this section we determine the rate of an $e^{+}e^{-}$ pair annihilation
in the presence of a third particle that carries away momentum and
enables production of only one photon. That particle can be an electron
or a positron. This process occurs in a positronium ion (Sec.~\ref{subsec:Ion})
and in the molecule $\text{Ps}_{2}$ (Sec.~\ref{subsec:Ps2Single}).
We confirm previously published results for both systems. This section
demonstrates our approach to computing annihilation amplitudes and
tests it in a three- and four-constituent systems. It prepares the
ground for the calculation of processes in which four particles in the
initial state interact, presented in Sec.~\ref{sec:Two-photon-annihilation}
and \ref{sec:ee-annihilation}.

\subsection{Single photon decay of the positronium ion $\text{Ps}^{-}\to e^{-}\gamma$\label{subsec:Ion}}

\begin{figure}[h]
\includegraphics[scale=0.35]{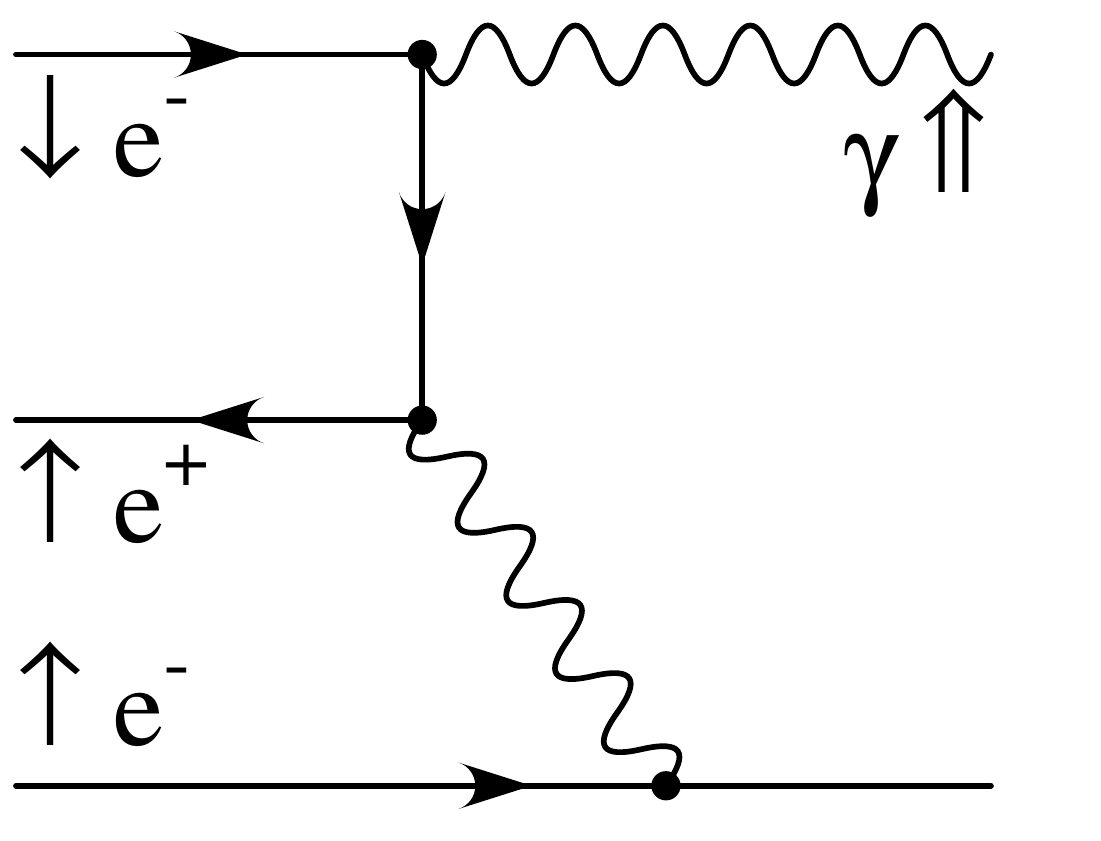}\hspace*{5mm}\includegraphics[scale=0.35]{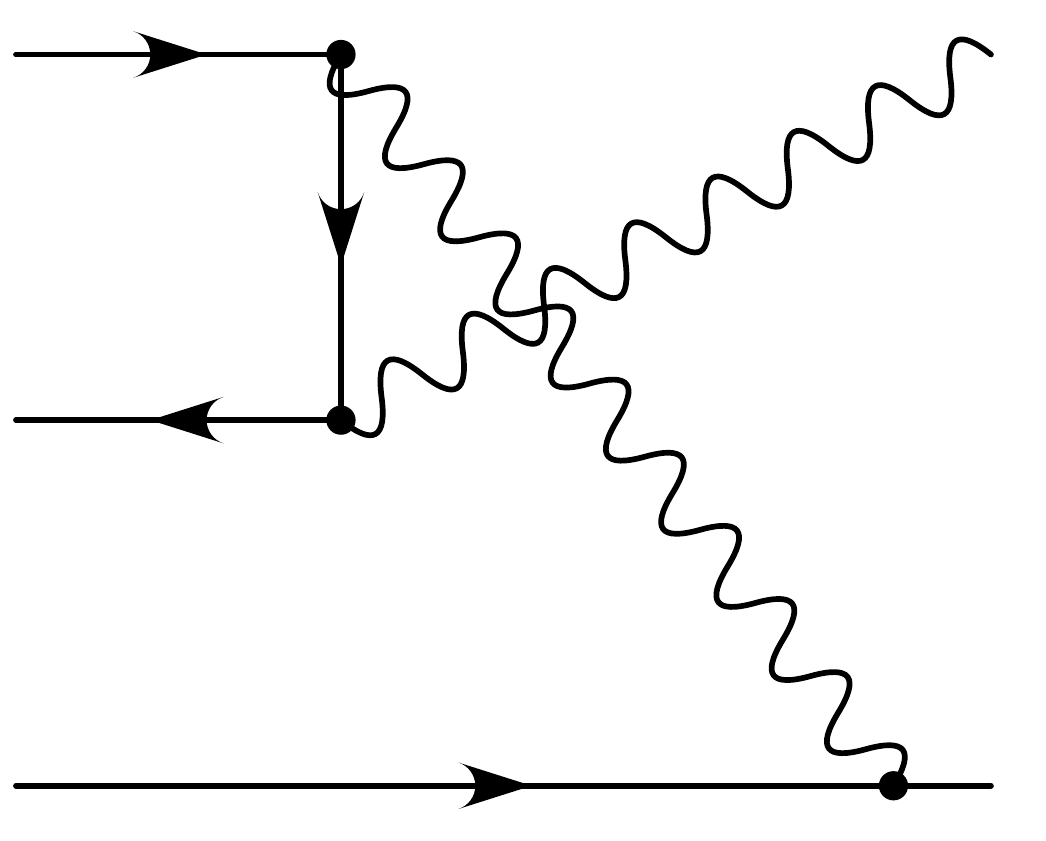}\hspace*{5mm}\includegraphics[scale=0.35]{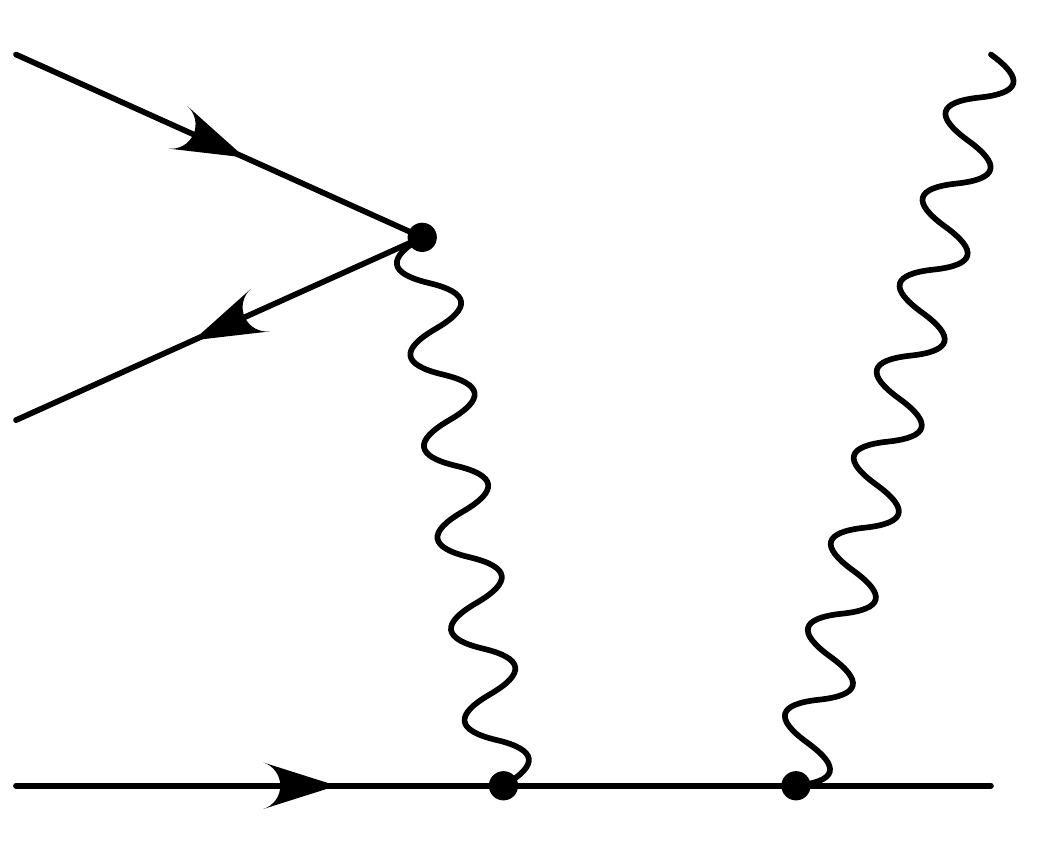}\hspace*{5mm}\includegraphics[scale=0.35]{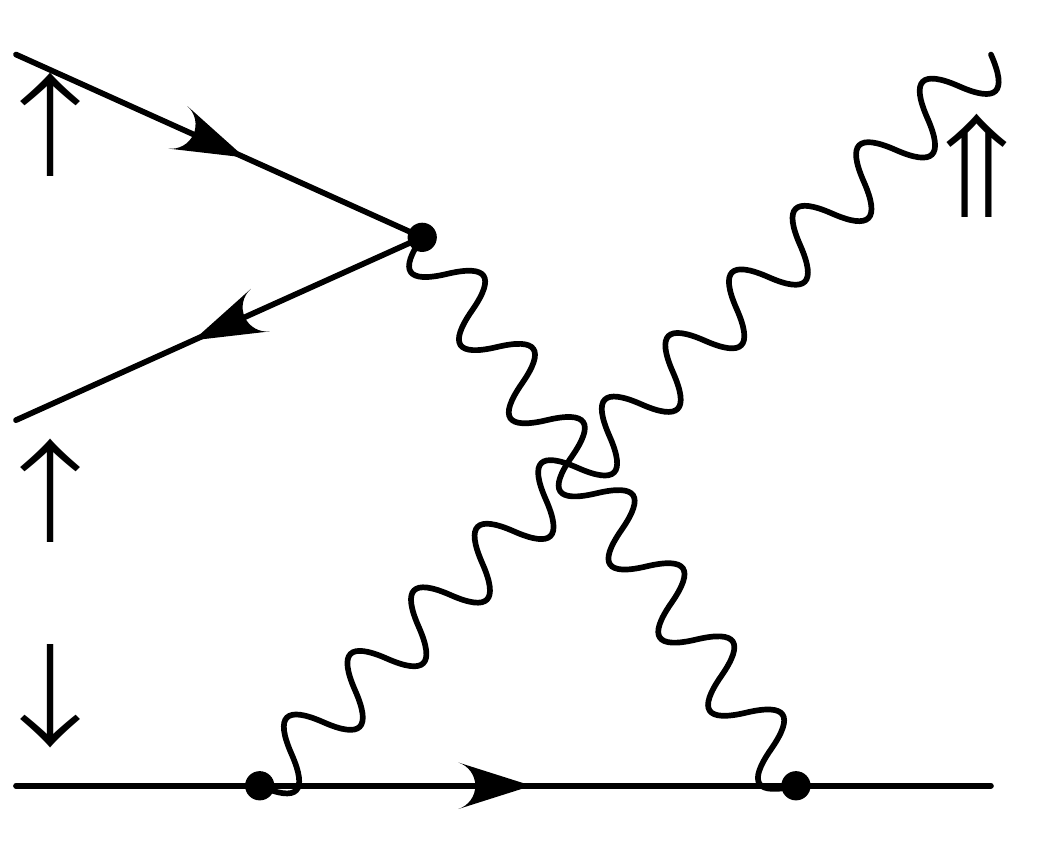}

\hspace*{15mm}(a)\hspace*{37mm}(b)\hspace*{37mm}(c)\hspace*{37mm}(d)

\caption{Examples of contributions to the single-photon decay of the positronium
ion, $\text{Ps}^{-}\to e^{-}\gamma$. Arrows indicate the spin projections
on the $z$ axis: plain arrows denote spins 1/2 and a double arrow
denotes spin 1 of the photon. Four more diagrams are obtained by switching
the roles of the two electrons $e^{-}$. In all Figures, time flows
horizontally from left to right.\label{fig:PsIon}}
\end{figure}

The positronium ion consists of two electrons and a positron. In the
vast majority of its decays, the positron encounters an electron with
which it forms a spin singlet and annihilates into two photons. ({In
the ground state of $\text{Ps}^{-}$ the electrons are in the spin-singlet
state, $\frac{\uparrow\downarrow-\downarrow\uparrow}{\sqrt{2}}$.
The positron can form a spin singlet or triplet with one of the electrons.
In the latter case, the annihilation produces at least three photons
and is much slower.)} However, there is also a rare decay channel
into a single photon,{{} $\text{Ps}^{-}\to e^{-}\gamma$}.
It can happen either when the two-photon annihilation is followed
by the absorption of one photon by the spectator electron, as shown
in {Fig.~\ref{fig:PsIon}(a,b); or by a single-photon
annihilation of a spin-triplet pair, with the photon scattering off
the spectator electron, as in Fig.~\ref{fig:PsIon}(c,d).}

At leading order in QED, this decay proceeds through
a total of eight Feynman diagrams. In addition to the four diagrams
shown in Fig.~\ref{fig:PsIon}, there are four in which the roles
of the two electrons are interchanged. 

Since the positronium ion is very weakly bound,
we  consider the annihilating particles to be at rest, similarly to
the standard analysis of the positronium atom annihilation \cite{Peskin:1995ev}.
We define the $z$-axis along the spin of the positron.

Since the ion $\text{Ps}^{-}$ is a spin 1/2 system, its decay amplitude
is fully characterized by two complex parameters \cite{Feynman:1963uxaV3}.
For example, we can choose the probability amplitudes of the photon
emission along the spin as one parameter, and in the opposite direction
as the other. In fact, it is sufficient to calculate one of them:
if the photon is emitted along the initial spin direction, it must
be right-handed, because of the angular momentum conservation: in
this case, the electron emitted in the opposite direction is also
right handed so the total projection of the spin on the $z$ axis
is 1/2 as in the initial state. When the photon is emitted in the
opposite direction, it must be left handed by the same argument. The
two amplitudes do not interfere because they describe distinct states
(right versus left handed particles). Because of parity conservation
in QED, probabilities of observing photons of each handedness must
be equal. Thus if we are interested only in the decay rate, it is
sufficient to calculate one of the amplitudes and multiply the resulting
rate by 2. 

We shall assume that a right handed photon and a
right handed electron are produced. Their momenta are along the positive
and negative $z$ axis, respectively.

In Fig.~\ref{fig:PsIon}, in panels (a,b) the annihilating
$e^{+}e^{-}$ pair is a spin singlet, and in (c,d) it is a spin triplet.
Since we choose the $z$ axis along the positron spin, that spin is
always up. It is immediately clear that amplitudes (a) and (d) vanish.
In (a), the annihilating electron forms a spin singlet with the positron,
so that electron's spin points down. But such an electron cannot emit
a photon whose spin points up. In (d), the annihilation occurs in
a spin triplet, so the spectator electron's spin is initially down;
again, it cannot emit the needed photon. Thus in our approach only
two amplitudes require an evaluation. (Similar considerations will
simplify our calculation of the decay $\text{Ps}_{2}\to\gamma\gamma$
described in Sec.~\ref{sec:Two-photon-annihilation}.)

Since we neglect the motion of initial state particles,
their spinors take a simple form. We write products of spinors on
each of the two fermion lines as a combination of Dirac gamma matrices
(cf.~Appendix \ref{sec:App1}). Multiplying these spinor combinations
by QED expressions for vertices and propagators, the total amplitude
for all eight diagrams becomes
\begin{align}
\mathcal{M}\left(e^{+}e^{-}e^{-}\rightarrow e^{-}\gamma\right)_{\text{free}} & =\frac{1}{\sqrt{2}}\left(\mathcal{M}_{e_{\uparrow}^{-}e_{\uparrow}^{+}e_{\downarrow}^{-}}-\mathcal{M}_{e_{\downarrow}^{-}e_{\uparrow}^{+}e_{\uparrow}^{-}}\right)=\left[\frac{2}{\sqrt{3}}-\left(-\frac{2}{\sqrt{3}}\right)\right]\frac{e^{3}}{4\me^{3}}=\frac{\left(4\pi\alpha\right)^{3/2}}{\sqrt{3}\me^{3}},\label{eq:2}
\end{align}
where $\alpha=\frac{e^{2}}{4\pi}$ (we use such units that $\epsilon_{0},$
$\hbar$, and $c$ are 1). 

The free-particle amplitude in Eq. (\ref{eq:2})
is related to the case of bound particles by (\cite{Peskin:1995ev},
page 149)
\begin{align}
\mathcal{M}\left(e^{+}e^{-}e^{-}\rightarrow\gamma+e^{-}\right)_{\text{bound}} & =\Psi\left(0,0,0\right)\mathcal{M}\left(e^{+}e^{-}e^{-}\rightarrow e^{-}\gamma\right)_{\text{free}},\label{eq:3}
\end{align}
where $\Psi\left(0,0,0\right)$ is the probability amplitude of all
constituents of the ion being at the origin. Its absolute value squared
is the expectation value of a product of two-particle delta functions
\cite{Frolov2007},
\begin{equation}
\left|\Psi\left(0,0,0\right)\right|^{2}=\left\langle
  \delta^{3}\left(\vr_{e^{+}e^{-}}\right)\delta^{3}\left(\vr_{e^{-}e^{-}}\right)\right\rangle
\equiv\left\langle \delta_{+--}\right\rangle a_{B}^{-6}\simeq 3.589\cdot10^{-5}\alpha^{6}m^{6},
\end{equation}
where $a_{B}=1/\left(\alpha\me\right)$ is the Bohr radius. In the
computation of the decay rate this expectation value must be divided
by $2!$ for the two identical electrons in the initial state [see Eq.~(\ref{Appen20})].
Another factor arises from the integration over the direction of the
photon emission: the probability of positron's spin projection on
an axis with the polar angle $\theta$ is $\cos^{2}\left(\theta/2\right)$
whose average is $1/2$. Remembering factor 2 accounting for both
photon polarizations we find the decay rate,
\begin{align}
\Gamma\left(\text{Ps}^{-}\rightarrow e^{-}\gamma\right) & =\frac{\left\langle \delta_{+--}\right\rangle }{2!}\cdot\frac{1}{2}\cdot2\cdot\frac{1}{9\pi}\cdot\left[\frac{\left(4\pi\alpha\right)^{3/2}}{\sqrt{3}\me^{3}}\right]^{2}\alpha^{6}m^{6}\cdot2m\label{eq:5}\\
 & =\frac{64}{27}\left\langle \delta_{+--}\right\rangle
   \pi^{2}\alpha^{9}m=0.0382 \text{ s}^{-1}.\label{eq:6}
\end{align}
Factor ${1}/{9\pi}$ results from the two-body phase space with
momentum ${4m}/{3}$ in the center-of-mass frame. The last factor
in (\ref{eq:5}), $2m$, comes from the electron spinor in the final
state. Our result agrees with Ref.~\cite{Kryuchkov_1994} which has
0.0392 using an older value of $\left\langle \delta_{+--}\right\rangle $
\cite{Mills:1981zzc}. 

\subsection{Single photon decay of the molecule {$\text{Ps}_{2}\to e^{+}e^{-}\gamma$
\label{subsec:Ps2Single}}}

Any three of the four constituents of $\text{Ps}_{2}$ can give rise
to the process analogous to {Fig.~\ref{fig:PsIon},}
possibly with the non-annihilating electron replaced by a positron.
This doubles the rate (not the amplitude: the non-annihilating particle
participating in the hard process becomes fast so the two processes
are distinguishable and do not interfere \cite{Frolov:2009qi}). There
is however an additional symmetry factor $1/2!$ due to identical
positrons, as discussed in Appendix \ref{sec:App3}. 

The ground state wave function of $\text{Ps}_{2}$
is symmetric in space (to minimize the kinetic energy). Since it must
be anti-symmetric in both electron and positron pairs, both electron
and positron pairs form spin singlets. The spin wave function is
\begin{equation}
\chi_{s}=\frac{e_{\uparrow}^{-}e_{\downarrow}^{-}-e_{\downarrow}^{-}e_{\uparrow}^{-}}{\sqrt{2}}\cdot\frac{e_{\uparrow}^{+}e_{\downarrow}^{+}-e_{\downarrow}^{+}e_{\uparrow}^{+}}{\sqrt{2}}.\label{eq:7}
\end{equation}
An extra factor 2 in the amplitude from the two ways of assigning
the positron role (annihilating or not) is partially cancelled by
$1/\sqrt{2}$ in the positron spin wave function{.
}In total, the numerical coefficient is twice that in the $\text{Ps}^{-}$
decay rate, 
\begin{align}
\Gamma\left(\text{Ps}_{2}\rightarrow e^{+}e^{-}\gamma\right) & =2\cdot\frac{\left\langle \delta_{+--}\right\rangle _{\text{Ps}_{2}}}{\left(2!\right)^{2}}\cdot\frac{1}{2}\cdot2\cdot\frac{1}{9\pi}\cdot\left[\sqrt{2}\frac{\left(4\pi\alpha\right)^{3/2}}{\sqrt{3}\me^{3}}\right]^{2}\alpha^{6}m^{6}\cdot2m\\
 & =\frac{128}{27}\left\langle \delta_{+--}\right\rangle _{\text{Ps}_{2}}\pi^{2}\alpha^{9}m,
\end{align}
in agreement with \cite{Frolov:2009qi}. 

\section{{Two-photon annihilation of $\text{Ps}_{2}$ \label{sec:Two-photon-annihilation}}}
\begin{figure}[h]
\includegraphics[scale=0.45]{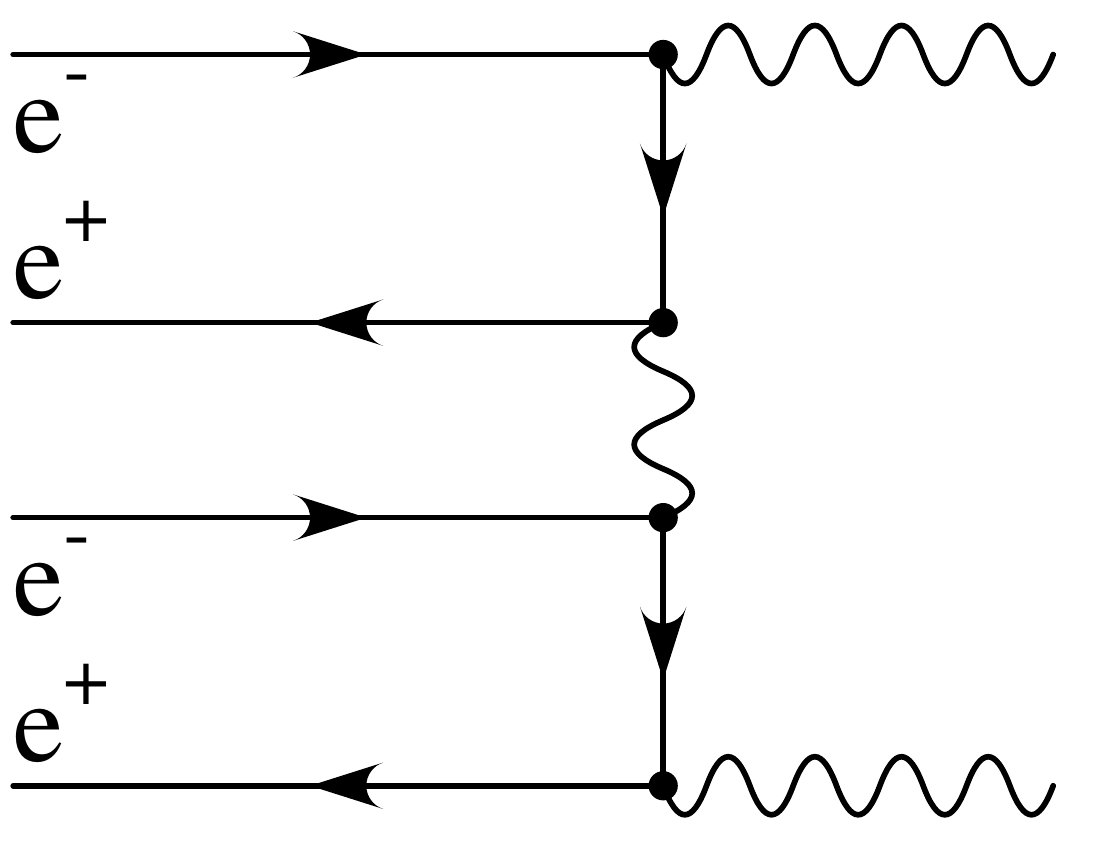}{\hspace*{8mm}}\includegraphics[scale=0.45]{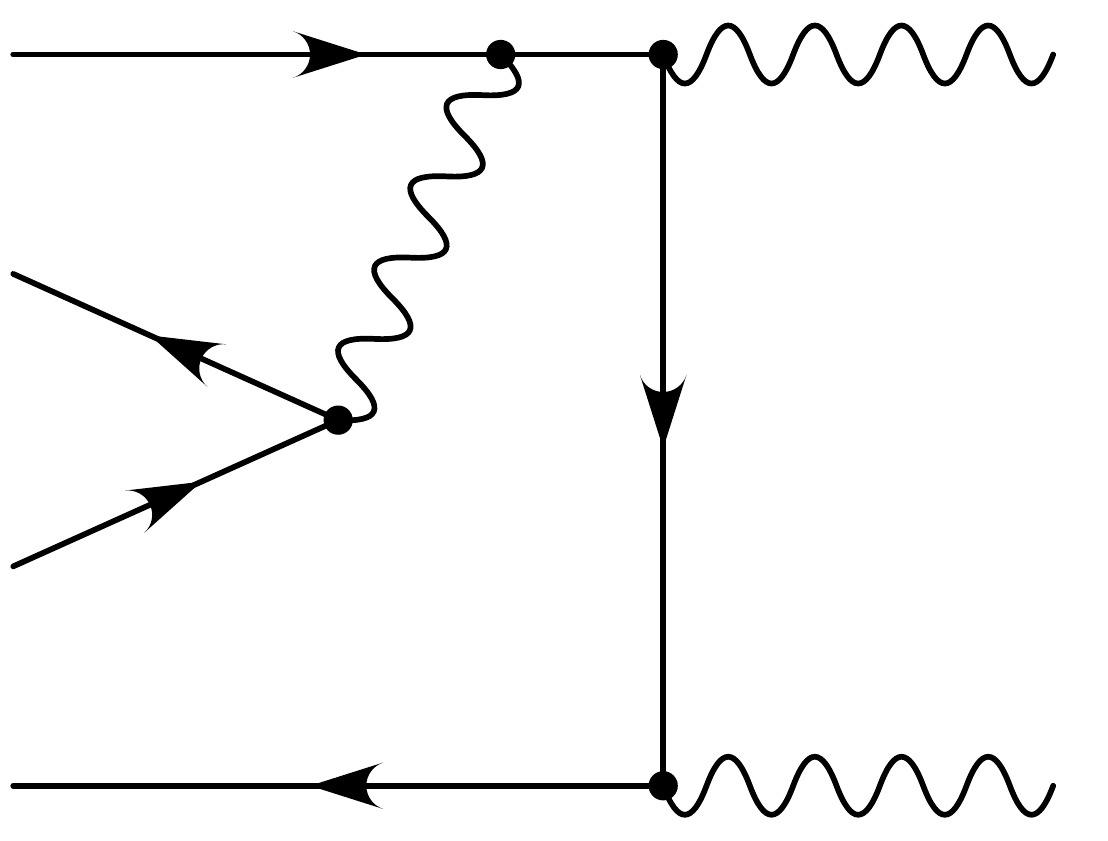}{\hspace*{8mm}}\includegraphics[scale=0.45]{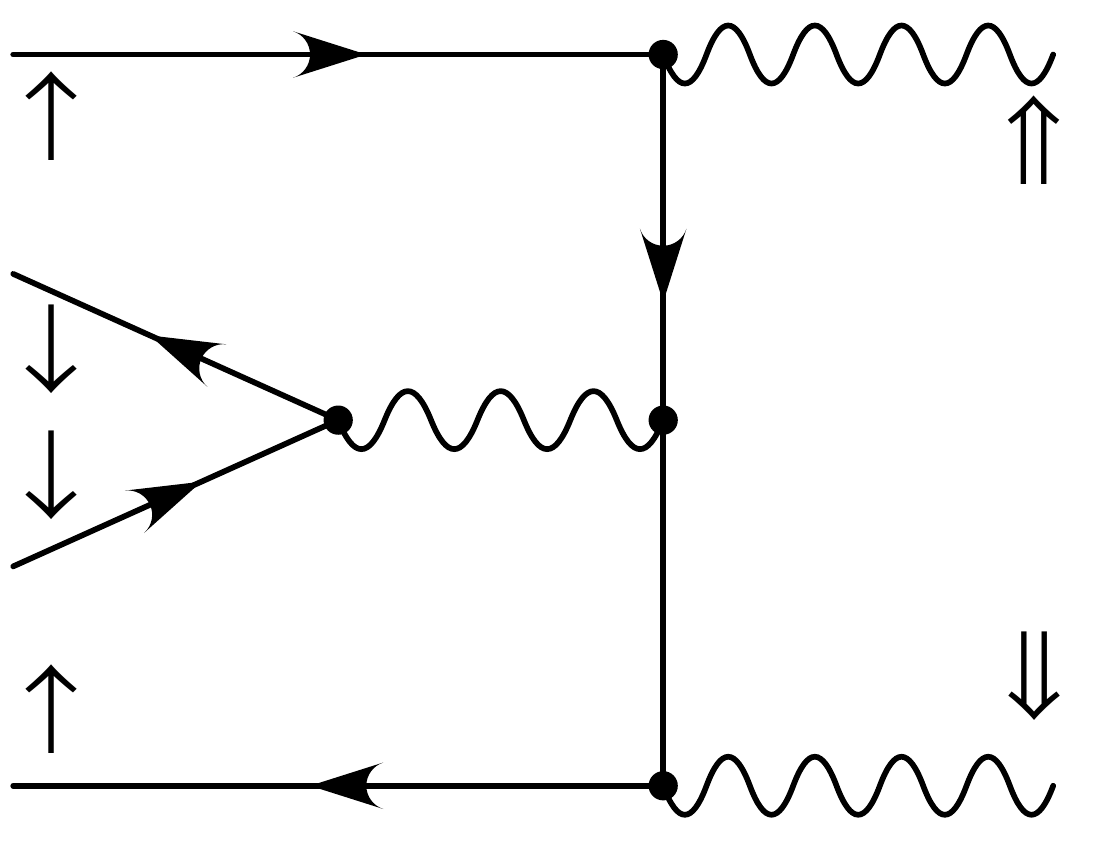}

{\vspace*{5mm}\hspace*{0mm}A\hspace*{55mm}B\hspace*{55mm}C}
\centering{}{\caption{Three types of contributions to the two-photon
    decay $\text{Ps}_{2}\rightarrow\gamma\gamma$.  Diagrams of type C
    turn out not to contribute to $\text{Ps}_{2}$ decays. One example
    of external particles' spins is shown: a spin-up fermion cannot
    emit a spin-down photon.\label{fig:dipositronium}}
}
\end{figure}

The two-photon annihilation of the molecule, $\text{Ps}_{2}\to\gamma\gamma$,
is a rare process in which both $e^{+}e^{-}$ pairs annihilate. Examples
of contributing diagrams are shown in Fig.~\ref{fig:dipositronium}. 

In this section we recalculate the rate of $\text{Ps}_{2}\to\gamma\gamma$
using explicit electron and positron spinors, as explained in Sec.~\ref{sec:Single-photon-decay}.
Since our result differs from that of Ref.~\cite{Perez-Rios:2014qia},
we describe our calculation and selected intermediate results in some
detail. 

We choose the $z$ axis along the momentum of final
state photons. We assume the photons are right handed (they must have
equal helicity because the initial state has zero angular momentum),
and at the end multiply the decay rate by 2 to account for left handed
photons. 

Fig.~\ref{fig:dipositronium} shows three types
of diagrams, A, B, and C. In A-type diagrams, both annihilating pairs
contribute a photon to the final state. There are $2^{4}=16$ diagrams
of this type: the order of photon vertices can be reversed of each
fermion line (factor $2^{2}$), the electrons can be assigned to either
annihilating pair (factor 2), and the final-state photons can be interchanged
(factor 2). 

In B-type diagrams, both photons in the final state
are emitted from the annihilation of single $e^{+}e^{-}$ pair. The
virtual photon resulting from the other $e^{+}e^{-}$ is absorbed
by the electron or the positron before annihilation. There are also
16 diagrams of this type: the photon can be absorbed by $e^{-}$ or
$e^{+}$ (factor 2); electrons can be selected in two ways for the
pair producing final-state photons, and so can positrons ($2^{2}$);
and again the final-state photons can be interchanged $\left(2\right)$.

The situation for the C-type diagrams is similar
to that of the B-type except, in this case, the photon emitted from the
triplet $e^{+}e^{-}$ pair is absorbed by the virtual electron. Interchanging
electrons, positrons, and real photons gives 8 C-type diagrams.
Due to the spin configuration of $\text{Ps}_{2}$, Eq.~(\ref{eq:7}),
C-type diagrams do not contribute. In case of aligned spins of the
$e^+ e^-$ pair annihilating into a single photon,  the remaining $e^{+}$
and $e^{-}$ must also have aligned spins and cannot emit two photons
with opposite spins, as shown in the last panel of
Fig.~\ref{fig:dipositronium}.
If the spins of the $e^+ e^-$ pair are opposite, C-type diagrams vanish
identically but we do not have an intuitive interpretation.

Following the strategy of the calculation of $\text{Ps}^{-}\to e^{-}\gamma$,
and using the expressions of spinors in terms of gamma matrices given
in Appendix \ref{sec:App1}, we find the matrix element for free particles
annihilating at rest,
\begin{align}
\text{A} & =-\frac{1}{2}\cdot\frac{ie^{4}}{\me^{4}},\quad\text{B}=\frac{1}{4}\cdot\frac{ie^{4}}{\me^{4}},\quad\text{C}=0,\\
\calM_{\gamma\gamma} & =\text{A}+\text{B}+\text{C}=-\frac{ie^{4}}{4\me^{4}}.
\end{align}
Accounting for the spin wave function in Eq.~(\ref{eq:7}) we find
the amplitude for the free-particle case,
\begin{align}
\mathcal{M}\left(e^{+}e^{-}e^{+}e^{-}\rightarrow\gamma_{R}\gamma_{R}\right)_{\text{free}} & =\frac{1}{2}\left(\mathcal{M}_{e_{\uparrow}^{-}e_{\uparrow}^{+}e_{\downarrow}^{-}e_{\downarrow}^{+}}+\mathcal{M}_{e_{\downarrow}^{-}e_{\downarrow}^{+}e_{\uparrow}^{-}e_{\uparrow}^{+}}-\mathcal{M}_{e_{\uparrow}^{-}e_{\downarrow}^{+}e_{\downarrow}^{-}e_{\uparrow}^{+}}-\mathcal{M}_{e_{\downarrow}^{-}e_{\uparrow}^{+}e_{\uparrow}^{-}e_{\downarrow}^{+}}\right)\label{eq:8a}\\
 & =\frac{\left(4\pi\alpha\right)^{2}}{2m^{4}}.\label{eq:8}
\end{align}
The photon momentum is $2m$ so the phase space gives $\frac{1}{8\pi}\cdot\frac{1}{2}$
where we have accounted for identical bosons in the final state. Since
the initial state has zero angular momentum, the distribution of photons
is isotropic. Remembering again about both photon helicities we find
the decay rate of the bound state,
\begin{align}
\Gamma\left(\text{Ps}_{2}\rightarrow\gamma\gamma\right) & =2\cdot\frac{1}{16\pi}\cdot\left[\frac{\left(4\pi\alpha\right)^{2}}{2m^{4}}\right]^{2}\frac{\left\langle \delta_{++--}\right\rangle \alpha^{9}m^{9}}{\left(2!\right)^{2}}\nonumber \\
 & =2\pi^{3}\alpha^{13}\left\langle \delta_{++--}\right\rangle m.\label{eq:12}
\end{align}
Using $\left\langle \delta_{++--}\right\rangle =4.5614\cdot10^{-6}$
from \cite{Frolov:2009qi}, confirmed in Ref.~\cite{Perez-Rios:2014qia},
\begin{equation}
\Gamma\left(\text{Ps}_{2}\rightarrow\gamma\gamma\right)=3.65\cdot10^{-11}{\text{
    s}^{-1}}.\label{eq:14}
\end{equation}
Instead of our coefficient 2 in Eq.~(\ref{eq:12}), {Ref.~\cite{Perez-Rios:2014qia}
has $521/1024$. As a result, they underestimate the rate of $\text{Ps}_{2}\rightarrow\gamma\gamma$
 by the factor 
\begin{equation}
2048/521=3.93.\label{ratio_gg}
\end{equation}
}\textbf{ }

\section{{Annihilation of $\text{Ps}_{2}$ into $e^{+}e^{-}$
\label{sec:ee-annihilation}}}

\begin{figure}[h]
\centering \includegraphics[scale=0.45]{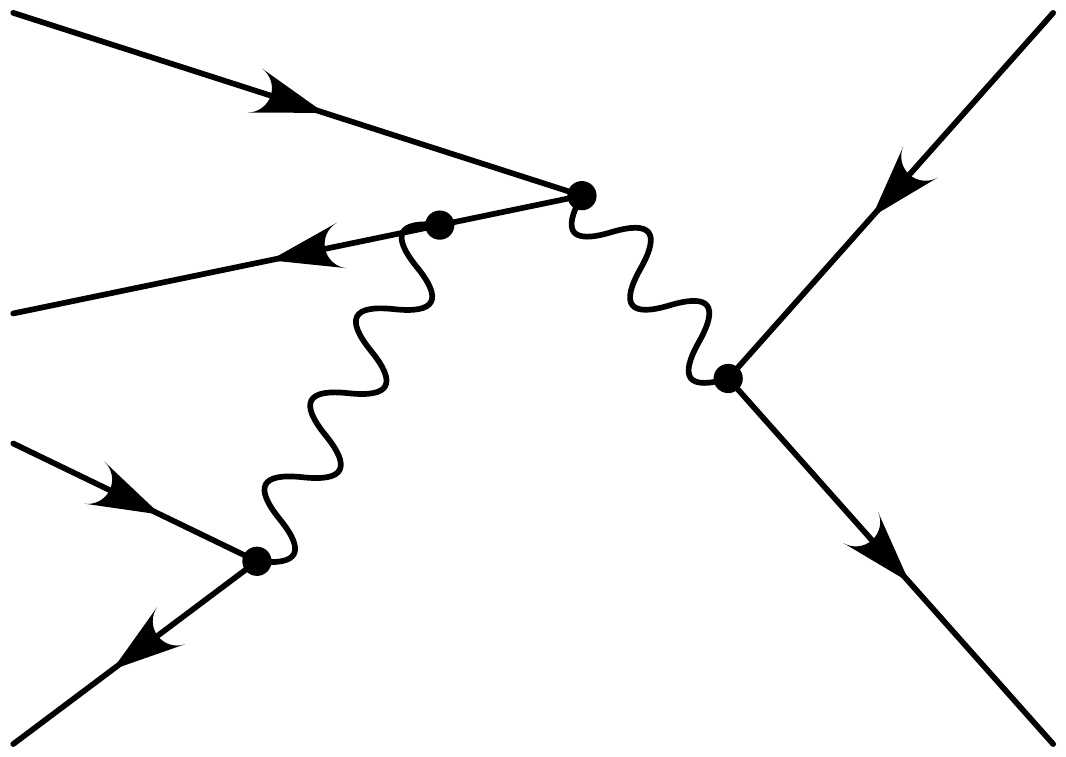}{\hspace*{12mm}}\includegraphics[scale=0.45]{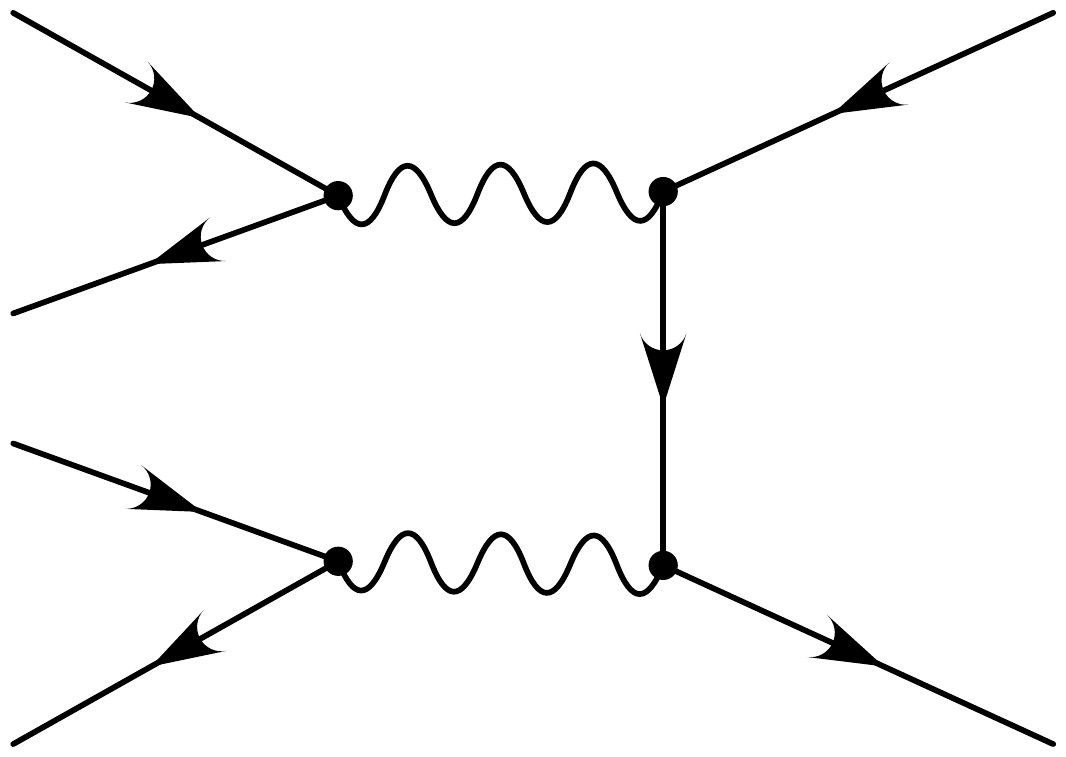}\\ I \hspace*{55mm} A \\[8mm] \includegraphics[scale=0.45]{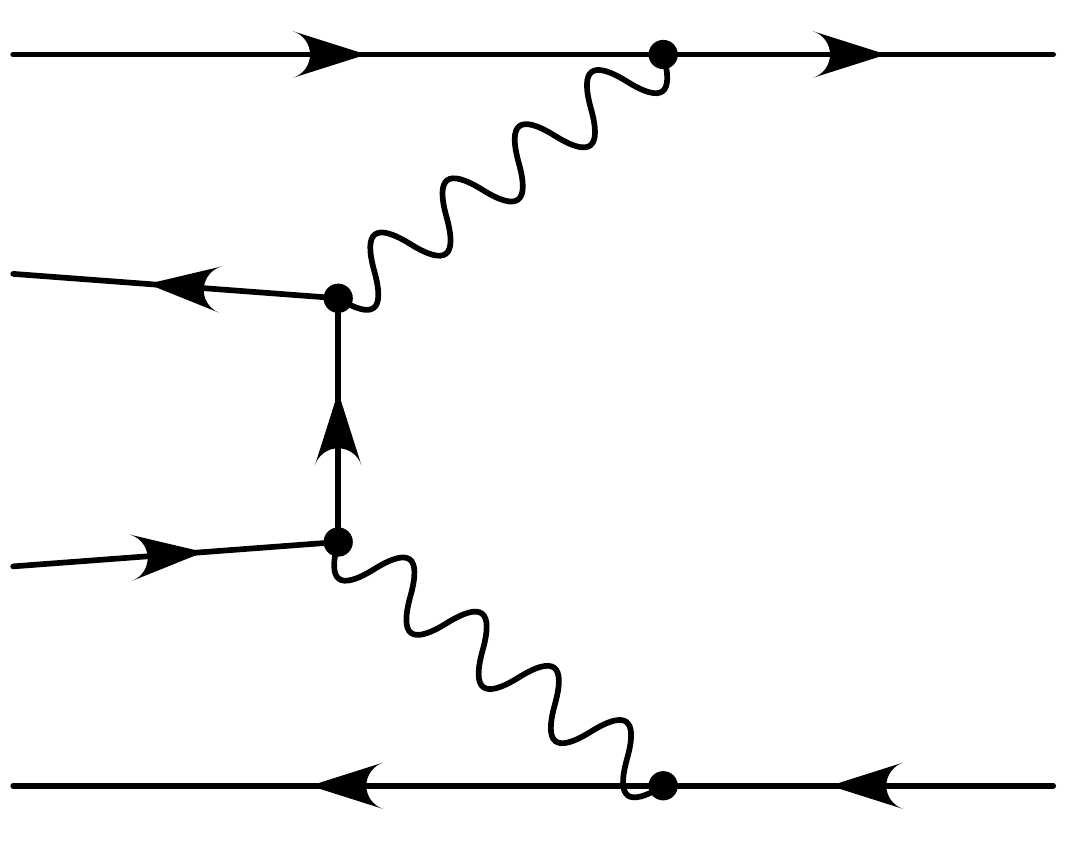}{\hspace*{12mm}}\includegraphics[scale=0.45]{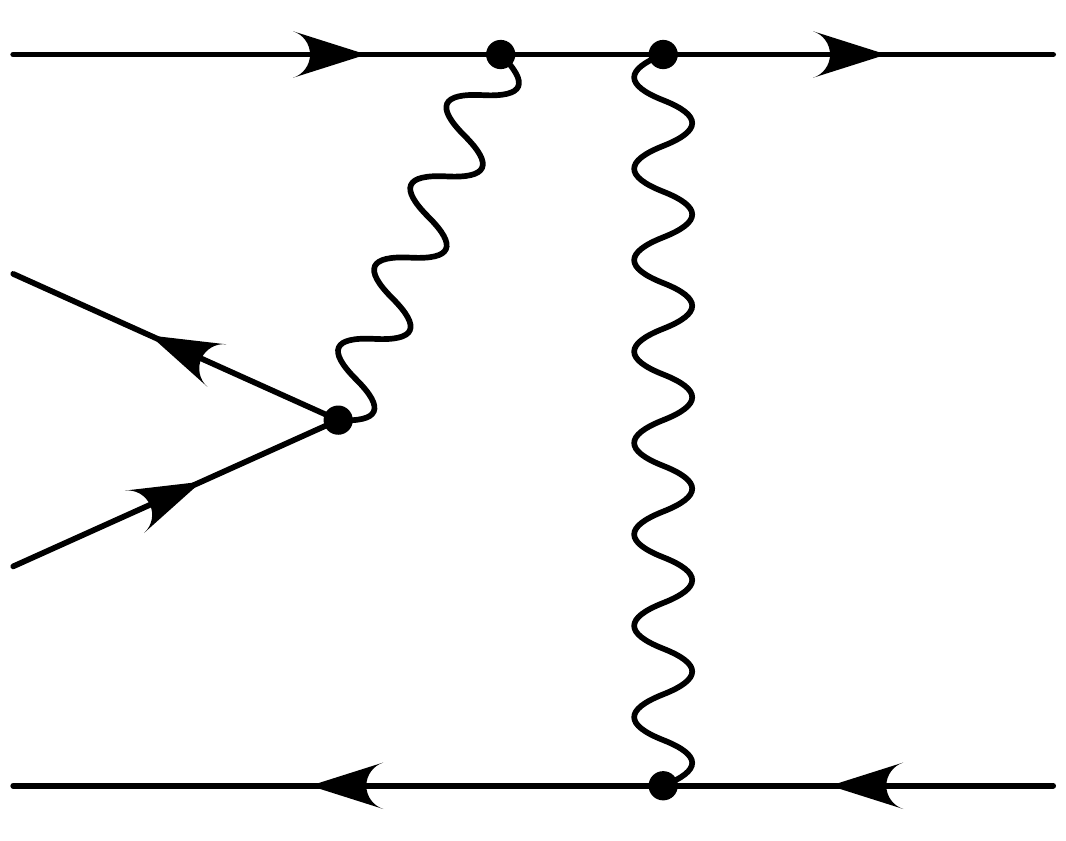}\\ B \hspace*{55mm} C 
\centering{}{\caption{Four groups of contributions to the annihilation $\text{Ps}_2\to e^{+}e^{-}$.
In group I, a single virtual photon produces the final state; such
diagrams do not contribute to the decay of the molecule $\text{Ps}_{2}$,
which has angular momentum 0. \label{fig:ee}}
}
\end{figure}

Fig.~\ref{fig:ee} shows examples of four types of diagrams contributing
to the annihilation $\text{Ps}_2\to e^{+}e^{-}$. More diagrams
are generated by changing the order of photon vertices wherever more
than one photon couples to an electron-positron line. In group C there
are also diagrams where the lower positron line absorbs the photon
resulting from the annihilation.

To evaluate their contributions to the decay of
$\text{Ps}_{2}$, we work in the rest frame of the molecule. We choose
the $z$ axis along the outgoing electron's momentum and assume that
it has spin up with respect to that axis. The positron must then have
spin down. We calculate the rate for this final state and double the
result to account for the opposite spin configuration.

Summing all photon orderings in each of the groups A, B, C in Fig.~\ref{fig:ee}
we find the amplitudes,
\begin{align}
\text{A} & =-\frac{1}{16}\cdot\frac{i\sqrt{3}e^{4}}{\me^{5}},\quad\text{B}=\frac{1}{8}\cdot\frac{i\sqrt{3}e^{4}}{\me^{5}},\quad\text{C}=-\frac{1}{4}\cdot\frac{i\sqrt{3}e^{4}}{\me^{5}},\\
\calM\left(e^{+}e^{-}e^{+}e^{-}\rightarrow e_{R}^{+}e_{R}^{-}\right)_{\text{free}} & =\text{A}+\text{B}+\text{C}=-\frac{3}{16}\cdot\frac{i\sqrt{3}e^{4}}{\me^{5}}.
\end{align}
The decay rate is found similarly to Eq.~(\ref{eq:12}). Phase space
is determined by electron's momentum, $\sqrt{3}m$, and gives a factor
$\frac{\sqrt{3}}{16\pi}$. There are two factors $2m$ for the final
state fermions. In the matrix element there is a net factor 2 just
like in Eq.~(\ref{eq:8a}). Together these factors give
\begin{align}
\Gamma\left(\text{Ps}_{2}\to e^{+}e^{-}\right) & =2\cdot\frac{\sqrt{3}}{16\pi}\cdot\left[\frac{2\cdot2}{\sqrt{2}^{2}}\frac{3\sqrt{3}\left(4\pi\alpha\right)^{2}}{16\me^{5}}\right]^{2}\frac{\left\langle \delta_{++--}\right\rangle \alpha^{9}m^{9}}{\left(2!\right)^{2}}\left(2m\right)^{2}\\
 & =\frac{27\sqrt{3}\pi^{3}\alpha^{13}}{2} \left\langle
   \delta_{++--}\right\rangle m \simeq 4.27\cdot10^{-10} \text{ s}^{-1}.\label{rate_ee}
\end{align}
The value previously published is 
$\Gamma\left(\text{Ps}_{2}\to e^{+}e^{-};
\text{{Ref.\ \cite{Frolov:2009qi}}}\right)
=2.32\cdot10^{-9} \text{ s}^{-1}.$
That reference has 147 instead of our 27 in Eq.~(\ref{rate_ee})
and overestimates the rate by the factor 
\begin{equation}
147/27=5.44.\label{ratio_ee}
\end{equation}

\section{{Conclusion\label{sec:Conclusion}}}

The goal of this study has been to explain the large ratio of about
250 of previously published predictions for $\Gamma\left(\text{Ps}_{2}\to e^{+}e^{-}\right)$
and $\Gamma\left(\text{Ps}_{2}\to\gamma\gamma\right)$. We have
demonstrated in Eqs.~(\ref{ratio_gg}) and (\ref{ratio_ee}) that previous
studies overestimated the first rate by 5.44 and underestimated the
second one by 3.93. Correcting for these factors we find the ratio
of $250/\left(5.44\cdot3.93\right)=11.7$.

Why do our results differ from previous studies? We believe that in
those works the spin wave function of $\text{Ps}_{2}$ was not properly
taken into account. In the case of $\text{Ps}_{2}\to e^{+}e^{-}$,
Ref.~\cite{Frolov:1995zz} says that the squared decay amplitude
is $\sum_{s_{5}s_{6}}4\left|M_{s_{5}s_{6}\uparrow\downarrow\uparrow\downarrow}\right|$
(we think there a trivial typo there: the square is missing) where
$s_{5,6}$ are the spins of the daughter electron and positron, and
arrows indicate spins of the positrons and electrons in the initial
state. We also fix the initial spin configuration as $\uparrow\downarrow\uparrow\downarrow$
(while computing amplitudes; we do eventually account for the full
spin wave function of $\text{Ps}_{2}$), but instead of summing over
all possible $s_{5,6}$, we take only $s_{5}=-s_{6}$ and sum over
the two values of $s_{5}$. The reason for this is that the total
spin projection of the final state must be zero, since the initial
state is a scalar. As a result, we find that the group of diagrams
I in Fig.~\ref{fig:ee} does not contribute, whereas Ref.~\cite{Frolov:1995zz}
states \emph{all of them contribute strongly to the result.} As a
result, Ref.~\cite{Frolov:1995zz} (and its subsequent refinements)
overestimates the rate by a factor of about 5.44. We also note that
by summing over all $s_{5,6}$, Ref.~\cite{Frolov:1995zz} includes
some contributions from triplet configurations of initial electrons
(and positrons): the initial-state electron spin configuration $\uparrow\downarrow$
is a mixture of the singlet $\left(\uparrow\downarrow-\downarrow\uparrow\right)/\sqrt{2}$
and the triplet $\left(\uparrow\downarrow+\downarrow\uparrow\right)/\sqrt{2}$,
whereas $\text{Ps}_{2}$ contains only the singlet. 

Regarding $\text{Ps}_{2}\to\gamma\gamma$, Ref. \cite{Perez-Rios:2014qia}
seems to disregard the spin wave function of the initial state, averaging
over all possible initial spins and summing of the final-state spins
(Eq.~(2) in Ref.~\cite{Perez-Rios:2014qia}). Bizarrely, the large
ratio (see our Eq.~(\ref{eq:250})) of $e^{+}e^{-}$ and $\gamma\gamma$
rates is attributed to different numbers of photon-electron vertices
in both processes. It is clear from Figs.~\ref{fig:dipositronium}
and \ref{fig:ee} that the number of vertices is four in both processes. 

Indeed, both processes are of the same order in $\alpha$ and both
involve $n=2$ extra participants in comparison to the leading decay
$\text{Ps}_{2}\to e^{+}e^{-}\gamma\gamma$. The remaining factor 11.7
between the two rates can be attributed to the difference in momentum
carried by the final state particles. Electrons, being massive, carry
a smaller momentum. Note that the two-body phase space, although proportional
to the daughter particle momentum, is actually larger by $\sqrt{3}$
for the $e^{+}e^{-}$ channel than for the $\gamma\gamma$ channel
because in the latter case there is a factor $1/2$ for identical
bosons. The smaller momentum of the electrons results in smaller values
of some $t$-channel propagators (they are less negative than in the
$\gamma\gamma$ case). 

In summary, we believe that our calculational approach clarifies and
simplifies studies of polyelectron decays. For example, in the case
of the decay $\text{Ps}^{-}\to e^{-}\gamma$, {in
Ref.~\cite{Kryuchkov_1994} eight amplitudes were first formally
summed and their sum was squared, resulting in 64 terms. Their evaluation
was characterized as }{\emph{rather involved}}{{}
and demanding a computer algebra system. In our approach not only
can the calculation be done by hand, but also the mechanism of the
decay is transparent. }

\section*{Acknowledgements}

We thank Arkady Vainshtein for helpful discussions. M.J.A. and A.C.
gratefully acknowledge the hospitality of the Banff International
Research Station (BIRS) where parts of this work were done. This work
was supported by the Natural Sciences and Engineering Research Council
of Canada.

\appendix

\section{{Spinors used in matrix elements\label{sec:App1}}}

Assuming that initial state particles are at rest,
$p_{i}=(\me,0,0,0)$, electron and positron spinors are
\begin{equation}
u_{\uparrow}=\left(\begin{array}{c}
1\\
0\\
0\\
0
\end{array}\right),\quad u_{\downarrow}=\left(\begin{array}{c}
0\\
1\\
0\\
0
\end{array}\right),\quad v_{\uparrow}=\left(\begin{array}{c}
0\\
0\\
0\\
-1
\end{array}\right),\quad v_{\downarrow}=\left(\begin{array}{c}
0\\
0\\
1\\
0
\end{array}\right).\label{Appen-1}
\end{equation}
Their products yield four by four matrices that we express in terms
of combinations of Dirac matrices,
\begin{align}
u_{\uparrow}v_{\uparrow}^{\dagger}=-\frac{1+\gamma^{0}}{2}\frac{\gamma^{1}+i\gamma^{2}}{2}, &  & u_{\uparrow}v_{\downarrow}^{\dagger}= & \frac{1+\gamma^{0}}{2}\frac{\gamma^{5}+\gamma^{3}}{2},\nonumber \\
u_{\downarrow}v_{\downarrow}^{\dagger}=\phantom{-}\frac{1+\gamma^{0}}{2}\frac{\gamma^{1}-i\gamma^{2}}{2}, &  & u_{\downarrow}v_{\uparrow}^{\dagger}= & -\frac{1+\gamma^{0}}{2}\frac{\gamma^{5}-\gamma^{3}}{2}.\label{Appen2}
\end{align}
Four-momenta of the final state photon $\left(k_{1}\right)$
and electron $\left(k_{2}\right)$ are
\begin{equation}
k_{1}=\left(\frac{4}{3}\me,0,0,\frac{4}{3}\me\right),\quad k_{2}=\left(\frac{5}{3}\me,0,0,-\frac{4}{3}\me\right).\label{Appen3}
\end{equation}
The spinor of the final state electron in $\text{Ps}^{-}\to e^{-}\gamma$
is
\begin{align}
u_{\uparrow}^{\dagger}(k_{2}) & =\sqrt{\frac{4}{3}}\left(\begin{array}{cccc}
1 & 0 & -\frac{1}{2} & 0\end{array}\right).\label{Appen4}
\end{align}

\section{Symmetry Factors for $\text{Ps}^{-}$and $\text{Ps}_{2}${\label{sec:App3}}}

The leading Fock state of $\text{Ps}^{-}$, a bound state of two electrons
and a positron, is
\begin{equation}
\left|\text{Ps}^{-}\left(\vP\right)\right\rangle =\int\widetilde{dk_{1}}\widetilde{dk_{2}}\psi_{s_{1}s_{2}s_{3}}\left(\vk_{1},\vk_{2},\vP\right)a_{s_{1}}^{\dagger}\left(\vk_{1}\right)a_{s_{2}}^{\dagger}\left(\vk_{2}\right)b_{s_{3}}\left(\vP-\vk_{1}-\vk_{2}\right)\left|0\right\rangle ,\label{Appen7}
\end{equation}
where $\widetilde{dk_{i}}=\frac{d^{3}k_{i}}{\left(2\pi\right)^{3}}$
and $a_{s}^{\dagger}\left(\vk\right)\left(b_{s}\left(\vk\right)\right)$
creates an electron (positron) with momentum $\vk$ and spin projection
$s$ and $\vP$ is the total momentum of the ion. In Eq.~(\ref{Appen7})
and onwards, summation over repeated indices is understood. In the
$\text{Ps}^{-}$ ground state, the wave function factorizes,
\begin{equation}
\psi_{s_{1}s_{2}s_{3}}\left(\vk_{1},\vk_{2},\vP\right)=\psi\left(\vk_{1},\vk_{2},\vP\right)\chi_{s_{1}s_{2}s_{3}},\label{Appen12}
\end{equation}
where $\chi_{s_{1}s_{2}s_{3}}=\frac{1}{\sqrt{2}}\left(\uparrow\downarrow-\downarrow\uparrow\right)\uparrow$
is the spin wave function, anti-symmetric in the electron spins $s_{1}$
and $s_{2}$, and $\psi\left(\vk_{1},\vk_{2},\vP\right)$ is symmetric
in $\vk_{1}$ and $\vk_{2}$. The corresponding position space wave
function is
\begin{align}
\Phi\left(\vr_{1},\vr_{2},\vr_{3}\right) & =\int\widetilde{dk_{1}}\widetilde{dk_{2}}\exp\left[i\vk_{1}\cdot\vr_{1}+i\vk_{2}\cdot\vr_{2}+i\left(\vP-\vk_{1}-\vk_{2}\right)\cdot\vr_{3}\right]\psi\left(\vk_{1},\vk_{2},\vP\right)\label{Appen9}\\
 & =e^{i\vP\cdot\vr_{3}}\int\widetilde{dk_{1}}\widetilde{dk_{2}}\exp\left[i\vk_{1}\cdot\left(\vr_{1}-\vr_{3}\right)+i\vk_{2}\cdot\left(\vr_{2}-\vr_{3}\right)\right]\psi\left(\vk_{1},\vk_{2},\vP\right)\\
 & \equiv e^{i\vP\cdot\vr_{3}}\phi\left(\vrho_{1},\vrho_{2}\right),\label{Appen10}
\end{align}
where $\vrho_{1}=\vr_{1}-\vr_{3}$ and $\vrho_{2}=\vr_{2}-\vr_{3}$
are positions of the two electrons relative to the positron. The Jacobian
of this shift is 1 and 
\begin{equation}
\phi\left(\vrho_{1},\vrho_{2}\right)=\int\widetilde{dk_{1}}\widetilde{dk_{2}}\exp\left(i\vk_{1}\cdot\vrho_{1}+i\vk_{2}\cdot\vrho_{2}\right)\psi\left(\vk_{1},\vk_{2},\vP\right).\label{Appen11}
\end{equation}
The spatial wave function is normalized by the condition $\int d^{3}\vrho_{1}d^{3}\vrho_{2}\left|\phi\left(\vrho_{1},\vrho_{2}\right)\right|^{2}=1,$
giving $\int\widetilde{dk_{1}}\widetilde{dk_{2}}\left|\psi_{s_{1}s_{2}s_{3}}\left(\vk_{1},\vk_{2},\vP\right)\right|^{2}=1$.
To find the normalization of $\left|\text{Ps}^{-}\left(\vP\right)\right\rangle $
we use anti-commutation relations,
\begin{align}
\left\{ a_{s}\left(\vk\right),a_{s^{\prime}}^{\dagger}\left(\vk^{\prime}\right)\right\} =\left\{ b_{s}\left(\vk\right),b_{s^{\prime}}^{\dagger}\left(\vk^{\prime}\right)\right\}  & =\left(2\pi\right)^{3}\delta^{3}\left(\vk-\vk^{\prime}\right)\delta_{ss^{\prime}},\nonumber \\
\left\{ a_{s}\left(\vk\right),b_{s^{\prime}}^{\dagger}\left(\vk^{\prime}\right)\right\} =\left\{ a_{s}^{\phantom{\dagger}}\left(\vk\right),b_{s^{\prime}}^{\phantom{\dagger}}\left(\vk^{\prime}\right)\right\}  & =0,\label{Appen14}
\end{align}
and find, using the anti-symmetry $\psi_{s_{2}s_{1}s_{3}}^{\ast}\left(\vk_{2},\vk_{1},\vP\right)=-\psi_{s_{1}s_{2}s_{3}}^{\ast}\left(\vk_{1},\vk_{2},\vP\right)$,
\begin{align}
 & \left\langle \text{Ps}^{-}\left(\vP^{\prime}\right)\right.\left|\text{Ps}^{-}\left(\vP\right)\right\rangle \nonumber \\
 & =\int\prod_{i=1}^{2}\widetilde{dk_{i}}\widetilde{dk_{i}^{\prime}}\psi_{s_{1}s_{2}s_{3}}\left(\vk_{1},\vk_{2},\vP\right)\psi_{s_{1}^{\prime}s_{2}^{\prime}s_{3}^{\prime}}^{\ast}\left(\vk_{1}^{\prime},\vk_{2}^{\prime},\vP^{\prime}\right)\cdot\left(2\pi\right)^{9}\delta^{3}\left(\vP-\vP^{\prime}\right)\delta_{s_{3}^{\phantom{\prime}}s_{3}^{\prime}}\\
 & \qquad\cdot\left[\delta^{3}\left(\vk_{1}^{\prime}-\vk_{1}\right)\delta_{s_{1}^{\prime}s_{1}^{\phantom{\prime}}}\delta^{3}\left(\vk_{2}^{\prime}-\vk_{2}\right)\delta_{s_{2}^{\prime}s_{2}^{\phantom{\prime}}}-\delta^{3}\left(\vk_{1}^{\prime}-\vk_{2}\right)\delta_{s_{1}^{\prime}s_{2}^{\phantom{\prime}}}\delta^{3}\left(\vk_{2}^{\prime}-\vk_{1}\right)\delta_{s_{2}^{\prime}s_{1}^{\phantom{\prime}}}\right]\nonumber \\
 & =2\left(2\pi\right)^{3}\delta^{3}\left(\vP-\vP^{\prime}\right)\int\widetilde{dk_{1}}\widetilde{dk_{2}}\left|\psi_{s_{1}s_{2}s_{3}}\left(\vk_{1},\vk_{2},\vP\right)\right|^{2}=2\left(2\pi\right)^{3}\delta^{3}\left(\vP-\vP^{\prime}\right).\label{Appen20}
\end{align}
The factor $2$ is related to the indistinguishability of the two
electrons. We compensate for it while calculating the decay rate.

In the case of $\text{Ps}_{2}$, the spin wave function is anti-symmetric
in both electrons and positrons, hence repeating the above steps we
obtain a factor of $\left(2!\right)^{2}=4$ in the normalization;
we divide by it when calculating the decay rates of $\text{Ps}_{2}\to\gamma\gamma,e^{+}e^{-}.$


\end{document}